\documentclass[pra, twocolumn, superscriptaddress, noshowpacs]{revtex4-1}

\usepackage{setspace}
\usepackage{amsmath}
\usepackage{graphicx}
\usepackage{dsfont}

\def\be{\begin{equation}}
\def\ee{\end{equation}}
\def\bea{\begin{eqnarray}}
\def\eea{\end{eqnarray}}
\def\nn{\nonumber}

\def\haf{\mbox{Haf}}
\def\perm{\mbox{Perm}}

\def\Pr{\mbox{Pr}}

\newcommand{\upb}{Integrated Quantum Optics, Universit\"at Paderborn, Warburger Strasse 100, 33098 Paderborn, Germany}
\newcommand{\prague}{FNSPE, Czech Technical University in Prague, Br\^ehov\'{a} 7, 119 15, Praha 1, Czech Republic}

\begin{document}
\title{Gaussian Boson Sampling}

\author{Craig S. Hamilton} \email{hamilcra@fjfi.cvut.cz}
\affiliation{%
\prague
}%
\author{Regina Kruse}
\affiliation{\upb}
\author{Linda Sansoni}
\affiliation{\upb}
\author{Sonja Barkhofen}
\affiliation{\upb}
\author{Christine Silberhorn}
\affiliation{\upb}
\author{Igor Jex }
\affiliation{%
\prague
}%

\begin{abstract}
Boson Sampling has emerged as a tool to explore the advantages of quantum over classical computers as it does not require a universal control over the quantum system, which favours current photonic experimental platforms.
Here, we introduce Gaussian Boson Sampling, a classically hard-to-solve problem that uses squeezed states as a non-classical resource. We relate the probability to measure specific photon patterns from a general Gaussian state in the Fock basis to a matrix function called the \textit{hafnian}, which answers the last remaining question of sampling from Gaussian states. Based on this result, we design Gaussian Boson Sampling, a \#P hard problem, using squeezed states. This approach leads to a more efficient photonic boson sampler with significant advantages in generation probability and measurement time over currently existing protocols. 
\end{abstract}

\maketitle

\textit{Introduction }Boson Sampling has sparked the imagination of theorists and experimentalists since it was introduced by Aaronson and Arkhipov (AABS) \cite{Aaronson:2013p7598}. It demonstrates the power of quantum over classical computation and provides evidence against the Extended Church-Turing theorem, without the need for the full power of a universal quantum computer. In photonic Boson Sampling, $N$ single photon Fock states are launched into a $N^2$-mode interferometer. Due to bosonic statistics, the probability to measure a specific photon pattern at the output depends upon the permanent of a submatrix of the interferometer unitary. The permanent is in the \textsf{\#P} complexity class \cite{Valiant:1979p11225} therefore this distribution is difficult to sample from, unless certain computational complexity classes are equivalent, which would have serious consequences for complexity theory. 
After this theoretical advance, several experimental groups performed the first demonstrations \cite{Broome:2013p7136, Tillmann:2013p10461, Spring:2013p7137, cres13npo2}.
However, since perfectly deterministic sources of single photons are not available (although recently, proof-of-principle Boson Sampling experiments with quasi-deterministic sources have been demonstrated \cite{wang_multi-photon_2016, he_scalable_2016, loredo_boson_2017}), they made use of post-selected photon-pair states from probabilistic photon-pair sources (such as two mode squeezed states) to emulate the single photon input states. 
This Postselected Fock Boson Sampling (PFBS), heralding of $N$ single photons from $N$ probabilistic sources, has an intrinsic exponential cost when scaling to high photon numbers and so cannot efficiently solve the Boson Sampling problem.
Lund {\it et al.} \cite{Lund:2014p10967} improve the scaling of the generation probability by a factor of ${N^2\choose N}$ by placing a probabilistic source in each of the $N^2$ input modes, a protocol known as Scattershot Boson Sampling (SBS), which is in the same complexity class as AABS. The improved scaling allows the protocol to be sampled efficiently, however it comes at the cost of an increased sampling space, comprising all possible combinations of $N^2 \choose N$ output patterns with each of the $N^2 \choose N$ possible input patterns.  
Recently, another way to improve the generation probability for high photon numbers was proposed by \cite{Barkhofen:2017p13761}. It is interesting to note that all of these schemes make use of Gaussian states but discard their Gaussian nature, as only a specific number of (postselected or heralded) single photons are retained from the complete distribution and the squeezers are driven in a low gain regime (mean photon number $\langle n \rangle \ll 1$).
Therefore, from an experimental perspective, it is valuable to investigate the Boson Sampling scheme with Gaussian states, appreciating the full Gaussian nature of the input states, which has also applications for the simulation of molecular vibronic spectra \cite{Huh:2015p12195}. This means to lift the constraint on pure single photon input states and consider i.e. squeezed states with a higher gain ($\langle n \rangle \lesssim 1$). 
In addition to an experimental interest, the appreciation of the full Gaussian nature also implies a strong theoretical relevance. Is a Boson Sampling problem with Gaussian states without the need for heralding in the same complexity class as sampling from single photon input states? This question has not yet been answered in general. Only for the special case of sampling from a multimode thermal state, the sampling problem could be placed in $\textsf{BPP}^\textsf{NP}$ \cite{RahimiKeshari:2015p11006, Tamma:2014p11008}, which is not as hard as AABS.

In this Letter we answer this question of sampling photons from a general Gaussian state and develop a new protocol we call Gaussian Boson Sampling (GBS).
Here, we utilise Single Mode Squeezed States (SMSS) as our non-classical resource, which then enter a linear interferometer and sample the output patterns in the photon number basis.
We first derive a new theoretical result that shows the probability to measure a specific photon output distribution from a general Gaussian state can be written in terms of a matrix function, the {\it hafnian}. As the hafnian is in \textsf{\#P} complexity class, we show that our exact GBS protocol is in \textsf{\#P} and argue that an approximate sampling problem with errors is also in the same complexity class. Contrary to the existing protocols, where the sampling matrix is directly given by the unitary of the interferometer, here the sampling matrix absorbs both the action of the interferometer and the overall shape of the Gaussian input state. This means that we can use a coherent superposition of all $N$-photon patterns from the Gaussian input and do not need to herald an exact input pattern, opposed to the other protocols where both input and output patterns determine the sampling problem. These two observations loosen the requirement on having single photon Fock states at the input and we are able to retain higher order photon number contributions. 
However, from an experimental point of view it is not only important to increase the generation probability, but also to accumulate enough statistics to extract the probabilities associated with each element of the sampling space. To this aim, GBS both decreases the size of the sampling space by a factor of $N^2\choose N$, compared to SBS, and increases the generation probability. As such, our GBS protocol has significant experimental advantages and puts photonic Boson Sampling with large number of photons within the grasp of current technology.

\begin{figure}[t]
\includegraphics[trim={1.5cm 9cm 1.5cm 8cm},clip,width=0.9\columnwidth]{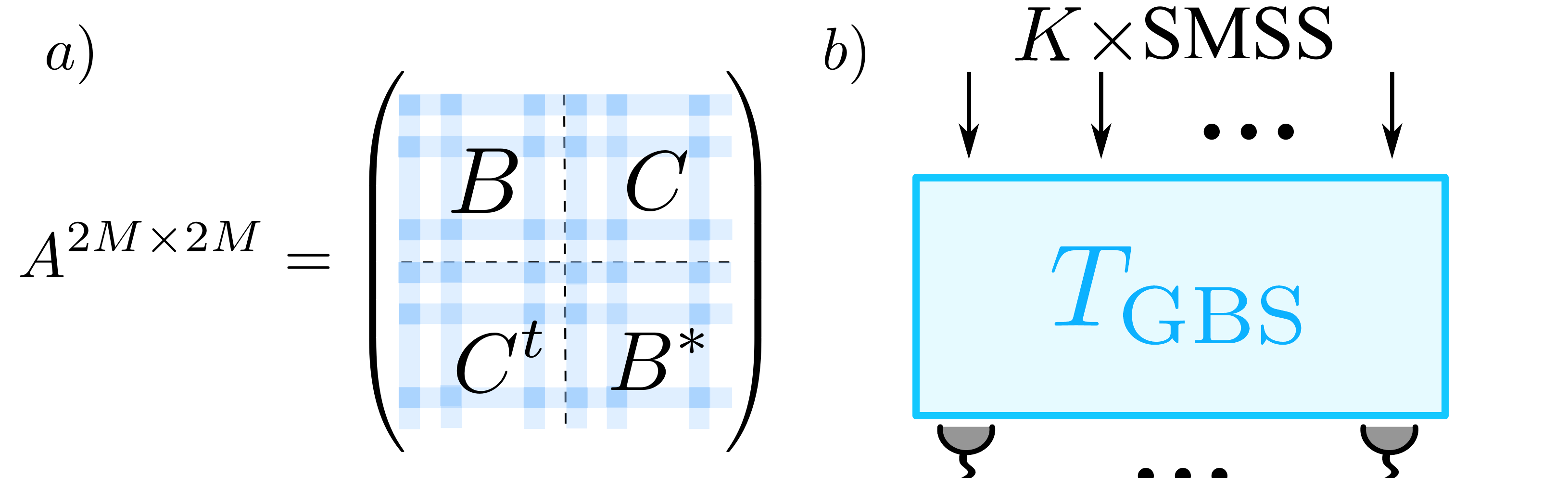}
\caption{a) Construction of submatrix $A_S$ from $A$, where highlighted rows/columns remain. Also shown is the structure of $A$($A_S$), which can be divided into 4 block matrices. b)$K$ SMSS enter a linear interferometer $T$ and at the output we measure the multimode squeezed state in the Fock state basis ($K\approx N \ll N^2=M$). The probability of a given pattern $\bar{n}$ is given by \eqref{eq:1}.}
\label{figure_rc_A}
\end{figure}

\textit{Photo-counts from a Gaussian state } 
Photonic Boson Sampling involves sending single photon Fock states into a linear interferometer, described by a matrix $T$, which transforms $M$ input modes into $M$ output modes. The probability of measuring a certain pattern of photons $\hat{\bar{n}} = \bigotimes^M_j n_j |n_j\rangle\langle n_j|$ ($n_j$ photons in output mode $j$) from $M$ modes of a quantum state $\hat{\rho}$ is $\Pr(\bar{n})=\mbox{Tr} [\hat{\rho}\,\hat{\bar{n}}]$. For Boson Sampling from Fock states $\Pr(\bar{n})$ depends upon the permanent of a matrix \cite{Scheel:2004p7177}
\be
\Pr(\bar{n}) =\frac{|\perm(T_S)|^2}{\bar{n}! \, \bar{m}!}\, ,
\label{AABS_eq}
\ee
where $\bar{m}$ is the input photon pattern, $\bar{n}!=n_1!n_2!..n_M!$ and $T_S$ is a submatrix of the linear transformation that depends upon where the photons enter and exit the interferometer. Here we derive a new expression for $\Pr(\bar{n})$ from a Gaussian state after passing an $M$-dimensional linear interferometer. This state is characterised solely by a $2M\times2M$ covariance matrix $\sigma$ and a displacement vector $d$ \cite{Ferraro:2005p4139},
\be
\sigma_{ij} = \frac{1}{2}\langle\{ \hat{\xi}_i, \hat{\xi}_j\}\rangle-d_i d_j, \quad d_j=\langle\hat{a}_j\rangle\nn\,,
\ee
where $\hat{\xi}$ run over all $\hat{a}^{}_j,\hat{a}^\dag_j$ (annihilation and creation operators for a photon in mode $j$) and we assume $d_j=0\,\,\forall j$. The details of this derivation are given in \cite{PRA}. Using phase space methods (similar to \cite{RahimiKeshari:2015p11006, Dodonov:1994p11102, Dodonov:1994p11080}), $\Pr(\bar{n})$ becomes the integral of the Q- and P-functions of the state and operator,
\be
\Pr(\bar{n}) = \pi^M 
\int \mathrm{d}^{2M}\boldsymbol{\alpha}\, Q_{\hat{\rho}}(\boldsymbol{\alpha}) P_{\bar{n}}(\boldsymbol{\alpha}).
\label{p_n_qp}
\ee
where 
$\mathrm{d}^{2M}\boldsymbol{\alpha}= \prod^M_{j=1} \mathrm{d}\alpha_j d\alpha_j^*$, $Q_{\hat{\rho}}$ is the Q-function of the state and $P_{\bar{n}}$ is the P-function corresponding to the operator $\hat{\bar{n}}$. This analysis leads to,
\be
\Pr(\bar{n}) = \left. \frac{1}{\bar{n}!\sqrt{|\sigma_Q|}} \prod_{j=1}^M \left(\frac{\partial^2}{\partial \alpha_j\partial \alpha^*_j}\right)^{n_j}\hspace{-2mm} e^{\frac{1}{2}\alpha^t_v A \alpha_v} \right|_{\alpha_v = 0}\, , 
\label{deriv_exp_fx}
\ee
where $\sigma_Q=\sigma+\mathds{I}_{2M}/2$, $\alpha_v=[\alpha_1,...,\alpha_M, \alpha_1^*,..,\alpha_M^*]$ and 
\be
A = \begin{pmatrix} 0_{} &\mathds{I}_M \\ \mathds{I}_M & 0_{}  \end{pmatrix} \left[\mathds{I}_{2M}-\sigma_Q^{-1} \right].
\ee
Note that $\sigma$ contains only the modes that are observed (i.e. measured). Any modes that are not observed are traced over to get a reduced covariance matrix. The sampling matrix $A$ can be divided into four block matrices, shown in figure~\ref{figure_rc_A}, which is a consequence of the initial structure of $\sigma$.
For simplicity we now focus on $n_j=\{0,1\}$ (we deal with $n_j\ge2$ in \cite{PRA}) for a total of $N$ photons and $2N$ derivatives (for $\partial \alpha_j, \partial \alpha_j^* $). The $N$ indices of the photons' mode-position are written in a vector $\mu$ of length $2N$ with entries $j$ and $j+M$ per photon. The $2N$ derivatives select the rows/columns of $A$ where the photons were measured; the other rows/columns will be discarded. This is illustrated in figure \ref{figure_rc_A}, where the intersection of the rows and columns where a photon was detected (highlighted in blue) form the entries of the submatrix $A_S$.
The expansion of the $2N$ derivatives leads to a summation over all perfect matching permutations (PMP) \cite{Callan:2009p12468, footnote2} of the vector $\mu$. For a general matrix $A_S$ this is
\be
\Pr(\bar{n})=\frac{1}{\bar{n}!\sqrt{ | \sigma_Q|}}\sum_{\mu'\in \mathrm{PMP}} \prod^N_{j=1} A_{S_{\mu'(2j-1), \mu'(2j)}}.
\ee
The sum over all PMP is exactly the hafnian of $A_S$, as defined by Caianiello \cite{Caianiello:1953p12510, Caianiello73}. Finally we arrive at
\be
\Pr(\bar{n}) =|\sigma|^{-1/2} \haf(A_S) / \bar{n}!
\label{eq:haf}
\ee
This new result relates the probability of a photon pattern $\bar{n}$ from a general Gaussian state to the hafnian of a matrix that characterises that state. This formula is applicable for any Gaussian state (i.e. any covariance matrix). We now use this result to develop a Boson Sampling protocol for Gaussian states, with squeezing contributions only ($B\neq0,\ C=0$ in figure \ref{figure_rc_A}).


\textit{Gaussian Boson Sampling with squeezed states }
As the hafnian is in the \textsf{\#P}-complete complexity class \cite{Valiant:1979p11225}, we can use Eq. (\ref{eq:haf}) to devise a quantum sampling problem akin to AABS. Whereas the permanent counts the (weighted) number of perfect matchings in a bipartite graph, the hafnian counts the number of perfect matchings in a general graph (not necessarily bipartite) \cite{Minc78}. Thus the hafnian is a more general function than the permanent, which is encapsulated in the formula, 
\be
\mbox{Perm}(G) = \mbox{Haf}\begin{pmatrix} 0 & G \\ G^t & 0 \end{pmatrix}\, ,
\label{perm_hf}
\ee
where $G$ is the graph's adjacency matrix.
 This means that any algorithm or black box that can accurately calculate the hafnian could also calculate the permanent, which is known to be \textsf{\#P}-hard even to approximate \cite{Aaronson:2013p7598}. Currently, there is no known algorithm to efficiently approximate the hafnian \cite{sankowski2003alternative, bjorklund2012counting}.\\
\indent We illustrate GBS with the scenario shown in figure~\ref{figure_rc_A} (b). $K\times$SMSS ($K\leq M$) enter an $M$-mode linear interferometer, described by a Haar random unitary $T$, with all modes being measured at the output. The squeezing transformation is described by,
\be
S =\begin{pmatrix} \bigoplus_{j=1}^{M}\cosh r_j & \bigoplus_{j=1}^{M} \sinh r_j \\ \bigoplus_{j=1}^{M} \sinh r_j &\bigoplus_{j=1}^{M}\cosh r_j \end{pmatrix}\, ,
\label{smat_GBS}
\ee
%
(4 block diagonal matrices \cite{footnote1}) and $r_j$ is the squeezing parameter of the $j^{th}$ mode, where at least $K$ of them are nonzero. The covariance matrix at the output of the interferometer is \cite{Simon:1994p4225},
\be
\sigma = \frac{1}{2} \begin{pmatrix}T & 0 \\ 0 & T^* \end{pmatrix} SS^\dag \begin{pmatrix}T^\dag & 0 \\ 0 & T^t \end{pmatrix},
\ee
and we arrive at ${A=B\oplus B^*}$, with,
\be
B = T \left(\oplus_{j=1}^{M}\tanh r_j \right )T^t\, .
\label{B_defn}
\ee
Using Eq. \eqref{eq:haf}, the probability to measure $\bar{n}$ (0 or 1 photon per mode) is then
\be
\Pr(\bar{n})=|\sigma_Q|^{-1/2}\left|\haf\left(B_S \right)\right|^2,
\label{eq:1}
\ee
where $B_{S}$ is the submatrix that comprises only the rows and columns where a photon was detected, i.e. the sampled output pattern. Note, that contrary to the sampling schemes from Fock states, we absorb the shape of the Gaussian input state into our sampling matrix $B$. Therefore, our scheme is independent on the exact location of the input photons and allows us to retain more than one photon per input mode. Nevertheless we have to ensure the complexity of the protocol, i.e. making $B$ complex enough. If we pump $K$ ($<M$) modes this means that $B$ in Eq. \eqref{B_defn} is a rank $K$ matrix. It is known that matrix rank determines the complexity of calculating the permanent \cite{Barvinok96}. Thus, we will assume a similar result for hafnians which means that to detect $N$ photons we will require to sample from (at least) a rank $N$ matrix. Therefore, we place a minimal requirement of $K=N$ SMSS at the input of our interferometer. 
%

%
\textit{Approximate GBS }%
Since a realistic Boson Sampler suffers from unavoidable error sources, we have to consider the problem of \textit{approximate} sampling. In AABS this problem corresponds to approximating the permanent up to additive error of matrices with random numbers from the complex normal distribution ($|\mathrm{GPE}|^2_\pm$) \cite{Aaronson:2013p7598}. AA show that this is in $\textsf{BPP}^{\textsf{NP}^{\mathcal{O}}}$, where $\mathcal{O}$ is an oracle for the AABS. Thus a fast classical algorithm for $\mathcal{O}$ would have severe consequences for the Polynomial Hierarchy. 
After this main result, AA introduce the \textit{permanent-of-gaussians} conjecture that expects approximate sampling with a \textit{multiplicative} error $\mathrm{GPE}_\times$ in \textsf{\#P}, and the \textit{permanent-anti-concentration} conjecture that surmises a polynomial-time equivalence of $|\mathrm{GPE}|_\pm^2$ and $\mathrm{GPE}_\times$. Provided that these two conjectures hold, then $\textsf{P}^\textsf{\#P}=\textsf{BPP}^\textsf{NP}$, meaning that approximate AABS has to be in \textsf{\#P} or the polynomial hierarchy collapses. Since the experimental implementations of AABS and GBS are similar they will suffer from the same error sources. In the following we use AAs main arguments of their hardness proof \cite{Aaronson:2013p7598} and transfer them to the GBS problem.

The main technical requirement that we have to fulfil is that the sampled submatrices $B_S$ in equation \eqref{eq:1} have random entries according to the complex normal distribution. Drawing the interferometer $T$ from the Haar measure and assuming all the squeezing parameters $r_j=r\, \forall \, j$, then $B$ simplifies to $B=\tanh r \times TT^t$, where $TT^t$ is from the Circular Orthogonal Ensemble (COE) of random matrices \cite{Mehta04}. An $N\times N$ submatrix of $B$ has random entries according to the complex normal distribution \cite{Jiang09COE}, if $N={O}(\sqrt{M})$. We can then use a $\textsf{BPP}^\textsf{NP}$ algorithm to find a particular COE matrix that includes $B_S$ as a submatrix. Based on this result, we know that an oracle $\mathcal{O}$ for GBS approximates the output probabilities up to \textit{additive} error using Stockmeyer's algorithm, i.e. $|\mathrm{GHE}|^2_{\pm}$, which means approximate GBS is in $\textsf{BPP}^{\textsf{NP}^\mathcal{O}}$. 
As in \cite{Aaronson:2013p7598}, we leave open the final proof that approximate GBS is in \textsf{\#P}. However, we give two conjectures for the hafnian that place approximate GBS into this complexity class. First, we formulate a \textit{hafnian-of-gaussians} conjecture, i.e. approximating the hafnian up to multiplicative error $\mathrm{GHE}_\times$ is in \textsf{\#P}. This is equivalent to AAs \textit{permanent-of-gaussians} conjecture. As the hafnian is a more general function than the permanent (see Eq. \eqref{perm_hf}), a hafnian approximation algorithm up to \textit{multiplicative} error would also approximate the permanent up to a multiplicative error, justifying our conjecture. 
Furthermore, we conjecture that the two approximations $|\mathrm{GHE}|^2_{\pm}$ and $\mathrm{GHE}_\times$ are polynomial-time equivalent, which we believe is justified due to the similar structure of the permanent and the hafnian. Provided these two conjectures hold, then $\textsf{P}^\textsf{\#P}=\textsf{BPP}^\textsf{NP}$, meaning that approximate GBS has to be in \textsf{\#P} or the polynomial hierarchy collapses.
%
%

\textit{GBS sampling patterns and generation probability }
Due to the nature of Gaussian states the total number of output photons is not fixed. This means that we have to sample all sets of output patterns containing $N$ photons in $M$ modes~\{$N\in [0, \infty)$ of size $C_N={M\choose N}$, assuming only 0 or 1 photon per output mode\}
\bea
\left\{\{p=|\sigma_Q|\}_0 ,\{p_1,p_2,..,p_{C_1}\}_1, ...,\{p_{1},p_{2},...,p_{C_N}\}_{N}, ...\right\} \nn\\
= \left \{ \{P_0\},\{ P_1\},\{ P_2\}, ...,\{ P_N\},..   \right \} \nn
\eea
where $p_{j}=\Pr(\bar{n})$ is the probability of a certain output pattern, given by Eq. \eqref{eq:1} and $\{P_N\}$ is the set of all output patterns with $N$ photons. Although we can retain more than one photon at the input of our interferometer, the restriction to either 0 or 1 photon per output mode means that we have to ensure low multiple photon events at the output. As in the original protocol by AA, this is guaranteed by the size of the interferometer, which leads for $N$ input photons in $N^2$ output modes to a mean photon number of $\tfrac{1}{N}$ per mode, satisfying our condition.
Since there exists no complexity proof for $N>\sqrt{M}$, we have to adapt the photon number generation of the SMSS to the dimension $M$ of the network to ensure the computational hardness of the problem. The probability to generate a total of $N$ Photon Pair Events (PPE, $2N$ photons) from $K\times\,$SMSS is given by the negative binomial distribution \cite{Hilbe11},
\begin{equation}
P_K(N)= {\frac{K}{2}+N-1 \choose N}\mathrm{sech}^{K}(r) \tanh^{2N}(r).
\label{neg_bin}
\end{equation}
The mean photon number of this distribution is $n_\mathrm{mean} = K\sinh^2(r)$ and the modal number (photon number with maximum probability) is $n_\mathrm{modal} = (K-1) \sinh^2(r)$. We can either operate in a regime where we focus on the probability of a specific photon number $N$ and choose the modal number $n_\mathrm{modal}=N$, or we consider a range of photon numbers $[N-c, N]$ (where c is a small integer) and set the mean photon number to $n_\mathrm{mean}=N$. Recalling our results from the previous section, we need at least $K\geq N$ SMSS at the input and an interferometer size of $M\geq N^2$ to saturate the complexity of an $N$-photon GBS experiment. In an experimental implementation we can choose one of these two regimes by fixing $K$ and adjusting the squeezing parameter $r$ accordingly.


\textit{Advantages of GBS }
To demonstrate the significant experimental advantages of our GBS protocol, we first compare the respective generation probabilities for fixed $N\times$ PPEs with existing protocols, which rely on probabilistic, post-selected PPEs from variable $K\times$ TMSS. Therefore, these types of protocols compelled to discard more than one photon pair from each squeezer. Thus, the probability to obtain $N$ single PPE from $K\times$ TMSS follows a binomial distribution
\be
P_\mathrm{prob}(N) = {K\choose N} \mbox{sech}^{2K}(r) \tanh^{2N}(r)\, .
\label{sbs_equ}
\ee
Comparing Eqs. \eqref{neg_bin} and \eqref{sbs_equ} for $K\times$TMSS and the same squeezing parameter $r$ for PFBS and GBS, we find that the ratio of these is
\bea
P_\mathrm{prob}(N) / P_\mathrm{GBS}(N) = {K \choose N}/{K+N-1 \choose N}\nn \\
\approx  \lim_{N\rightarrow \infty,K>N} \left (\frac{K-N}{K-1}\right)^N.
\eea
Comparing SBS, which uses $K=N^2$ TMSS (with $\bar{n}\approx1/N$), with GBS, we gain an $e$-fold increase in the probability to generate $N$ photons. Still, an additional advantage of GBS is that we only require a low number of squeezers, $K\approx N \ll N^2=M$ to saturate the complexity of an $N$-photon experiment. In this regime, GBS has significant experimental advantages over PFBS protocols, as the probability to generate useable photons scales exponentially better. Summarising, we gain a quadratic reduction in the number of required resources compared to SBS and an exponential increase in the generation probability compared to PFBS. In both cases, we gain an additional factor of two in the number of generated photons since we do not herald.

Yet, the main advantage of GBS does not lie in the increase of the generation probability, especially when compared to SBS. The biggest improvement of GBS lies in the reduction of the sampling space. In SBS, we have to sample the distribution of ${N^2 \choose N} \times {N^2 \choose N}$ possible combinations of input and output patterns, i.e. it is equivalent to randomly sample from ${N^2 \choose N}$ AABS problems (this combinatorial factor scales as $(eN)^N$ for large $N$). In GBS however, we do not have to condition on the exact location of the input photons, reducing our sampling space by a factor of $N^2 \choose N$, compared to SBS. 
This means that in an experiment we can take less data sets and save an exponential factor 
in the required measurement time. 


\textit{Conclusions} We have introduced Gaussian Boson Sampling, which uses the easy-to-achieve experimental resource of SMSS to implement a Boson Sampling problem. We derived a new expression for the output probabilities from a general Gaussian state and showed that they are related to a matrix function called the hafnian. Calculating the hafnian is a computationally hard problem in the complexity class \textsf{\#P} and we provided evidence that even approximating a GBS problem is difficult. Our result answers questions in previous work as to the complexity of Boson Sampling with Gaussian states \cite{Aaronson:2013p7598, RahimiKeshari:2015p11006, Huh:2015p12195}. Due to the symmetry of quantum mechanics, we can reverse the problem of GBS and use the same result to explain a Fock state input to an interferometer with Gaussian-basis measurements (see open problem 4 in \cite{Aaronson:2013p7598}). Within experimental quantum optics starting with a squeezed state, using linear optical transformations and postselected measurement outcomes is a very common method to create different families of photonic states. This means that GBS includes other photonic boson sampling protocols as special cases, which can be most readily seen from SBS, as we show in \cite{PRA}, but also includes other boson sampling problems such as those involving Sch\"{o}dinger cat states and photon added/subtracted states \cite{Rohde:2015p11547, Olson:2015p10963, Seshadreesan:2015p12695}. 
Let us note that this formalism allows us to handle the main source of noise in photonic systems, photon loss and dark-counts, in a very natural way as both are Gaussian operations. This makes lossy GBS easy to deal with compared to lossy PFBS \cite{Aaronson:2016p12815}, though it remains an open question how much loss we are able to tolerate and retain the \textsf{\#P} complexity of the scheme (as opposed to the $\textsf{BPP}^{\textsf{NP}}$ complexity of thermal states \cite{RahimiKeshari:2015p11006}.) 
Most importantly, we showed that a new class of photonic boson samplers can be realized by replacing the single photon inputs with squeezed light. GBS is a way to achieve a photonic boson sampler with a significant number of photons to demonstrate the supremacy of quantum computers. 

\textbf{Acknowledgements: }
C.S.H. and I. J. received support from the Grant Agency of the Czech Republic under grant No. GA\v{C}R 17-00844S  and from M\v{S}MT RVO 68407700. This work has received funding from the European Union's Horizon 2020 research and innovation program under the QUCHIP project Grant No.~641039. 

The authors would like to thank A. Arkhipov, A. Bj\"orklund, S. Rahimi-Keshari and T. C. Ralph for useful comments. 

\bibliography{GBS_Hamilton}

\begin{thebibliography}{36}
\bibitem [{foo({\natexlab{a}})}]{footnote2}%
  \BibitemOpen
  \href@noop {} {} ({\natexlab{a}}),\ \bibinfo {note} {pMP are permutations
  $\mu$ that obey 1) $\mu(2j-1)<\mu(2j)$ \& 2) $\mu(2j)<\mu(2(j+1)) \,\forall
  \,j$}\BibitemShut {NoStop}%
\end{thebibliography}

\end{document}